\documentclass[twocolumn,floatfix,amsmath,amssymb,prb,showpacs,superscriptaddress,hyperref]{revtex4-2}
\usepackage{graphicx}
\usepackage{color}
\usepackage{bbold}
\usepackage{tikz}
\usepackage{dcolumn}
\usepackage{bm,listings}
\usepackage{dsfont}
\usepackage{ulem}

\usepackage{amsfonts}
\usepackage{ulem}

\usepackage{comment}
\begin{document}

\title{Husimi dynamics generated by non-Hermitian Hamiltonians}
\date{\today}
\author{Katherine Holmes}\author{Wasim Rehman}\author{Simon Malzard}\author{Eva-Maria Graefe}
\address{Department of Mathematics, Imperial College London, London, SW7 2AZ, United Kingdom}

\begin{abstract}
The dynamics generated by non-Hermitian Hamiltonians are often less intuitive than those of conventional Hermitian systems. Even for models as simple as a complexified harmonic oscillator, the dynamics for generic initial states shows surprising features. Here we analyse the dynamics of the Husimi distribution in a semiclassical limit, illuminating the foundations of the full quantum evolution. The classical Husimi evolution is composed of two factors, (i) the initial Husimi distribution evaluated along phase-space trajectories, and (ii) the final value of the norm corresponding to each phase-space point. Both factors conspire to lead to intriguing dynamical behaviours. We demonstrate how the full quantum dynamics unfolds on top of the classical Husimi dynamics for two instructive examples. 
\end{abstract}

\maketitle

\textit{Introduction}---An underlying classical dynamics is central in the interpretation of quantum features. The comparison between quantum and classical behaviour in phase space is of particular value. Among the quantum phase-space distributions, the Husimi distribution stands out for being strictly positive and normalisable \cite{milburn1986quantum,altland2012quantum,Veronez_2013,veronez2016topological}. The quantum dynamics of the Husimi distribution is described by a partial differential equation of potentially infinite order. When truncated after the first order, in Hermitian systems it reduces to a Liouvillian classical phase-space flow, that provides the \textit{classical bones} for the \textit{quantum flesh} of the full dynamics generated by the higher order derivatives \cite{comment_Wigner}. 

In the present paper we consider the phase-space dynamics generated by non-Hermitian Hamiltonians, which have moved into the focus of intense research interest over the past decade (see, e.g., \cite{Moiseyev2011non,ashida2020non} and references therein). Complex energies provide simple heuristic models of losses or gains in quantum (and other wave) systems. Quantum dynamics generated by non-Hermitian Hamiltonians are also closely related to master equations appearing in descriptions of open quantum systems \cite{wisemanBook,daley2014quantum,graefebradley,klauck2019observation,graefenatphoton}. The special class of $PT$-symmetric Hamiltonians, describing systems with a certain balance of loss and gain, can lead to intriguing dynamical features at the intersection of conservative and dissipative behaviours \cite{roberts1992chaos,el2018non,christodoulides2018parity,longhi2018parity,bender2019pt}. We are only beginning to glimpse the rich structure of their quantum-classical correspondence. Phase-space considerations have yielded much new insight into static properties of non-Hermitian systems in comparison to their classical counterparts \cite{Keating_2006,Altmann_2013_Leaking,Hen_2010,Ketz_2018,Ketz_2021_intensity_stats,Joe2022}.

Non-Hermitian systems are often discussed on the backdrop of complexified classical equations of motion with complex position and momentum \cite{bender1999pt}. Alternatively, the dynamics of Gaussian states in a classical limit leads to dynamics on a real phase space with a changing metric structure \cite{Graefe2011}. For quadratic systems the two approaches have an elegant geometric connection \cite{bhattacharyya2022exponential,graefe2012complexified}. The quantum-classical correspondence of dynamics beyond the Gaussian approximation of the quantum state, however, has hitherto not been analysed, constituting a crucial gap in the understanding of non-Hermitian quantum systems. Here we provide a new perspective on these dynamics in terms of a classical phase-space evolution for arbitrary initial states.

We will show that for non-Hermitian Hamiltonians a semiclassical approximation leads to a Husimi evolution governed by two competing factors, the transported initial distribution along a set of phase-space trajectories, and a factor reflecting the change of the quantum norm as a function of the initial phase-space point, a \textit{norm landscape}. The interaction between these two components leads to nontrivial dynamics, already in example systems as simple as a harmonic oscillator with a complex frequency. We shall demonstrate the correspondence between the full quantum and the classical Husimi flows for two illustrative examples, a damped anharmonic oscillator with bistability, appearing in various physical applications \cite{drummond1980quantum,rigo1997quantum}, and an anharmonic oscillator with $PT$-symmetry. 

For simplicity we consider time-independent Hamiltonians. The ideas can be straight-forwardly generalised to Hamiltonians with explicit time-dependence. Throughout the paper we use scaled units with $\hbar=1$.

\textit{Husimi dynamics in the Harmonic oscillator}---
Let us begin with a brief summary of the dynamics in the standard quantum harmonic oscillator described by the Hamiltonian
\begin{equation}
\hat{H}=\omega(\hat{a}^{\dagger}\hat{a}+\frac{1}{2}\hat{I}), 
\label{eqn:H_HO}
\end{equation}
where $\omega\in\mathds{R}^{+}$ denotes the frequency of the oscillator. Coherent states
\begin{equation}
|z\rangle=\hat D(z)|0\rangle=e^{z\hat a^\dagger-z^*\hat a}|0\rangle,
\end{equation}
with 
\begin{equation}
z=\frac{1}{\sqrt{2}}(q+i p),   
\end{equation} 
and $p,q\in\mathds{R}$, 
remain coherent, with minimal uncertainty, under the dynamics generated by the Hamiltonian (\ref{eqn:H_HO}). An initial coherent state $|\psi(0)\rangle=|z_c\rangle$, localised at $z_c$, evolves as
\begin{equation}
|\psi(t)\rangle = e^{-i\frac{\omega}{2} t}|z_c(t)\rangle\quad{\rm with}\quad
    z_c(t)= z_c e^{-i\omega t},
\end{equation}
thus following the classical trajectories. 

The coherent states are an overcomplete basis of the Hilbert space, and their overlaps with a state $|\psi\rangle$ can be used to define the Husimi representation $Q(z)$ \cite{Drummond1981Husimi, milburn1986quantum} of the state as 
\begin{equation}\label{eqn:husimi_defn}
Q(z) = \langle z |\psi\rangle\langle\psi  |z\rangle.
\end{equation}
In the harmonic oscillator for an arbitrary initial state, an initial Husimi distribution is rotated rigidly in time with frequency $\omega$, just as a classical phase-space density would be.  

\textit{A complex harmonic oscillator}---
Let us now consider a complexified version of the harmonic oscillator described by the non-Hermitian Hamiltonian 
\begin{equation}
\hat{K}=(\omega-i\gamma)(\hat{a}^{\dagger}\hat{a}+\frac{1}{2}\hat{I}), 
\label{eqn:C.H.O.}
\end{equation}
where $\gamma, \omega \in \mathbb{R}^{+}$. Here $\gamma$ represents an effective damping. 
The eigenstates are the usual harmonic oscillator states $|n\rangle$, with corresponding complex eigenenergies $\lambda_n=(\omega-i\gamma)(n+\frac{1}{2})$.

Just as in the Hermitian limit, the complex harmonic oscillator leaves coherent states coherent \cite{graefe2010classical}. For an initial coherent state $|\psi(0)\rangle = |z_c\rangle$, the solution to the time-dependent Schr\"{o}dinger equation is given by
\begin{equation}
\label{eqn:dynpsiCS_cHO}
|\psi(t)\rangle = e^{-i\frac{\omega-i\gamma}{2}t}e^{-\frac{|z_c|^2}{2}(1-e^{-2\gamma t})}|z_c(t)\rangle,
\end{equation}
with
\begin{equation}
\label{eqn:cHOcs_zt}
    z_c(t)=z_c e^{-i(\omega-i\gamma)t}.
\end{equation} 

\begin{figure}[t]
\hspace{-0.312cm}
\includegraphics[width=0.5\textwidth]{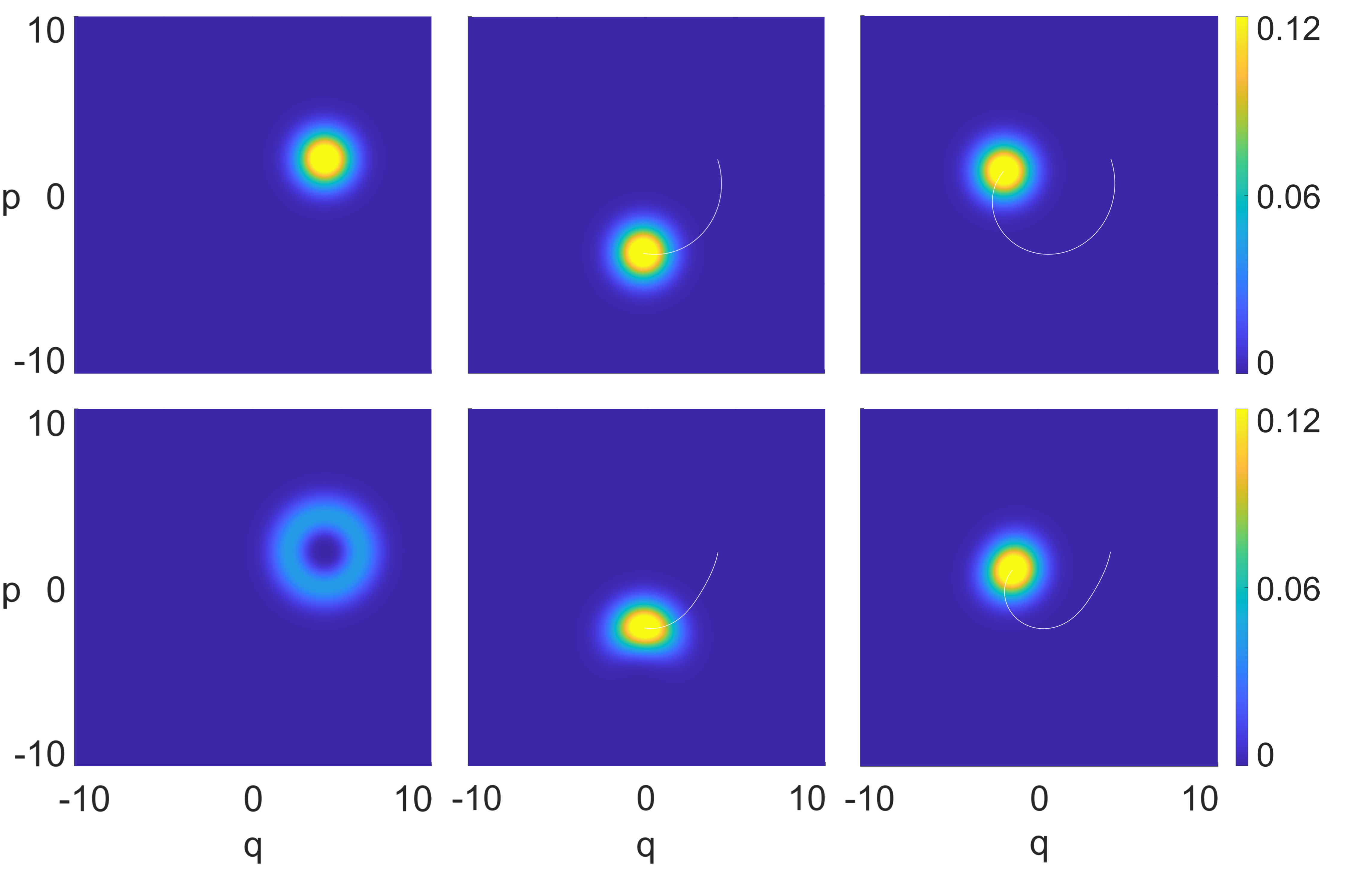}\hspace{-0.312cm}
\caption{\label{fig_nHermHO_Husimi_dyn1} Renormalised Husimi distributions at times $t=0$ (left), $t=\frac{2\pi}{3}$ (middle) and $t=\frac{4\pi}{3}$ (right) for a complex harmonic oscillator (\ref{eqn:C.H.O.}) with $\omega=1$ and $\gamma=0.15$. The top and bottom rows depict the renormalised Husimi distributions for an initial harmonic oscillator state $|n\rangle$ displaced to $z_c=\frac{1}{\sqrt{2}}(4+2i )$ for $n=0$ (top) and $n=2$ (bottom). The white lines depict the trajectories of the expectation values up to the specified times.}
\end{figure}

In contrast to the Hermitian case, the dynamics for more general initial states are not simply transported along the same trajectories as the coherent-state motion. 
From the direct solution of the time-dependent Schr\"odinger equation, we find the time-evolved Husimi function for an arbitrary initial state $|\psi(0)\rangle$ with Husimi distribution $Q_0(z)$ is given by
\begin{equation}
Q(z,t) = Q_{0}(\zeta_0(z,t))e^{-\gamma t}e^{-|z|^2(1-e^{-2\gamma t})},
\end{equation}
where $\zeta_0(z,t)$ denotes the functional dependence of the initial condition $\zeta_{0}$ of the trajectory
\begin{equation}
\label{eqn:cHO_Hamflow}
\zeta(t)=\zeta_0e^{-i(\omega+i\gamma)t},
\end{equation}
on $z,t$, such that $\zeta(t)=z$. That is, the time-dependent Husimi distribution consists of two factors; the initial distribution transported along the trajectories (\ref{eqn:cHO_Hamflow}), 
and a phase-space distribution 
\begin{equation}
\label{eqn:normls_cHO}
w(z,t):=e^{-\gamma t}e^{-|z|^2(1-e^{-2\gamma t})},
\end{equation}
 reflecting the growth/decay structure on phase space. Importantly, for a non-vanishing imaginary part, the trajectories (\ref{eqn:cHO_Hamflow}) do not agree with the trajectories of a coherent state given by (\ref{eqn:cHOcs_zt}). Instead, the interplay between the transported initial distribution and the distribution reflecting the different norm dynamics across phase space (\ref{eqn:normls_cHO}) returns the coherent-state trajectory for an initial coherent state.
 
 For an initial harmonic oscillator state $|n\rangle$ (which is not a minimum uncertainty state for $n\neq0$) displaced to $z_c$, with Husimi distribution
\begin{equation}
Q_{0}(z) = \frac{|z-z_c|^{2n}}{n!}\, e^{-|z-z_c|^2},
\end{equation}
the time-evolved Husimi function is given by
\begin{equation}
Q(z,t)=e^{-2\gamma(n+\frac{1}{2})t-|z-z_ce^{-i(\omega-i\gamma)t}|^2}\frac{|z-z_ce^{-i(\omega +i\gamma) t}|^{2n}}{n!}.
\end{equation} 
For $n=0$ this reduces to 
\begin{equation}
Q(z,t)=e^{-\gamma t}e^{-|z-z_c(t)|^2},
\end{equation}
with $z_c(t)=z_ce^{-i(\omega -i\gamma)t}$,
confirming the previous observation that a coherent state remains coherent under the dynamics, following the trajectory (\ref{eqn:cHOcs_zt}). For $n\neq 0$ the initial ring-shaped distribution transported along the outwards spiralling trajectories conspires with the Gaussian distribution that grows around the centre of phase space. In stark contrast to the Hermitian case, this leads to a deformation of the overall state following a different trajectory from the coherent state. This is illustrated in figure \ref{fig_nHermHO_Husimi_dyn1}, where snapshots of the Husimi distribution at different times for $n=0$ and $n=2$ are displayed.

\textit{Semiclassical dynamics of Husimi distributions}---
For general non-Hermitian systems with Hamiltonian $\hat K$, it follows from the Schr\"odinger equation that
\begin{align}
    i \frac{\partial{Q}}{\partial t}&= \langle z | \hat{K} | \psi \rangle \langle \psi | z \rangle - \langle z | \psi \rangle \langle \psi | \hat{K}^{\dagger} | z \rangle.\label{eqn:general_dQdt}
\end{align}
For simplicity, we assume that the Hamiltonian is an analytic function of creation and annihilation operators, expanded in a normal-ordered power series as 
\begin{equation}
\hat{K} = \sum_{m,n=0} K_{mn}\hat{a}^{\dagger m}\hat{a}^{n},
\label{eqn:genK}
\end{equation}
where $K_{mn}$ are complex coefficients. 
The equation of motion (\ref{eqn:general_dQdt}) of the Husimi distribution then takes the form
\begin{equation}
\begin{split}
i \frac{\partial Q}{\partial t} =\sum_{m,n=0}&K_{mn} z^{*m}\langle z | \hat{a}^{n}|\psi\rangle\langle\psi|z \rangle\\ &-K_{mn}^{*} z^{m}\langle z |\psi\rangle\langle\psi| \hat{a}^{\dagger n} |z \rangle,
\label{eqn:gen_kdot}
\end{split}
\end{equation}
where we have used that $\hat a^n|z\rangle=z^n|z\rangle$.
Further, using that \cite{Holm,milburn1986quantum} 
\begin{align}
\langle z | \psi\rangle\langle \psi|\hat{a}^{\dagger n} |z \rangle &= e^{-|z|^2}\frac{\partial^n}{\partial z^n}(Q(z) e^{|z|^2}) \label{eqn:identity1} \\
\langle z |  \hat{a}^n |\psi\rangle\langle\psi | z \rangle &= e^{-|z|^2}\frac{\partial^n}{\partial z^{*n}}(Q(z)e^{|z|^2}), \label{eqn:identity4}
\end{align}
we recognise equation (\ref{eqn:gen_kdot}) as a higher order differential equation for $Q(z,t)$. While $z$ and $z^*$ are regarded as independent variables here, for brevity, we omit the explicit dependence on $z^*$ in the notation. 

The classical  dynamics of the Husimi distribution, resulting from the leading orders in the semiclassical parameter $|z|^{-2}$ \cite{milburn1986quantum} is given by
\begin{equation}
\frac{\partial Q}{\partial t} + i\frac{\partial K}{\partial z}\frac{\partial Q}{\partial z^{*}} - i\frac{\partial K^{*}}{\partial z^{*}}\frac{\partial Q}{\partial z} -2\Gamma Q = 0. \label{eqn:husimi_pde_kstar-k}
\end{equation}
Here $K$ denotes the classical phase-space function 
\begin{equation}
K = \sum_{m,n=0} K_{mn}z^{* m}z^{n},
\end{equation}
and $\Gamma=\frac{1}{2i}(K-K^*)$ is the classical counterpart of the anti-Hermitian part of the Hamiltonian.

For a given initial Husimi distribution $Q_0(z)$, we solve equation (\ref{eqn:husimi_pde_kstar-k}) using the method of characteristics \cite{strauss2007partial} to find
\begin{align}
\label{eqn:Husimi_solution}
Q(z,t) & = Q_{0}(\zeta_0(z,t)) w(z,t),
\end{align}
where $\zeta_0(z,t)$ denotes the initial condition of the trajectories 
. 
\begin{equation} 
\label{eqn:char_traj}
\dot \zeta = -i\frac{\partial K^{*}}{\partial \zeta^{*}},
\end{equation}
such that $\zeta(t)=z$. In terms of real phase-space  coordinates $p$ and $q$ the equations of motion  (\ref{eqn:char_traj}) become
\begin{equation} 
\label{eqn:char_traj_pq}
\dot p = -\frac{\partial H}{\partial q}+\frac{\partial \Gamma}{\partial p},\ {\rm and}\ 
\dot q = \frac{\partial H}{\partial p}+\frac{\partial \Gamma}{\partial q},
\end{equation} 
where $H=\frac{1}{2}(K+K^*)$ denotes the classical counterpart of the Hermitian part of the Hamiltonian. The function $w(z,t)$ is given as the solution of the differential equation
\begin{equation}
\dot w(z,t)=2\Gamma(\zeta_0(z,t))w(z,t),
\end{equation}
with the initial condition $w(z,0)=1$. To obtain $w(z,t)$, every individual phase-space point is transported \textit{backwards} along a trajectory $\zeta(t)$ of (\ref{eqn:char_traj}), and the corresponding value of $w(z,t)$ changes by locally decaying/growing exponentially with a rate proportional to the current value of $\Gamma$ along the trajectory. This reflects the influence of the change of norm of the quantum system, and we will refer to the distribution of final values $w(z,t)$ at a given time as the \textit{norm landscape}. 

Equation (\ref{eqn:Husimi_solution}) is the central result of this paper. For a Hermitian system the time-evolved classical Husimi distribution is simply given by the initial Husimi distribution transported along the classical trajectories. In contrast to this, in the non-Hermitian case the time-evolved Husimi function consists of a product of the initial Husimi function transported along the trajectories $\zeta(t)$ (the solution of Hamilton's equations with Hamiltonian $K^*$ as opposed to $K$) weighted by the norm landscape at the given time. 

Equation (\ref{eqn:Husimi_solution}) can alternatively be derived from a coherent state approximation of the dynamics. The Husimi distribution of a time-evolved state $|\psi(t)\rangle=\hat U|\psi(0)\rangle$, with time evolution operator  $\hat{U}=e^{-i\hat{K}t}$, is by definition given by
\begin{equation}
\label{eqn:husimi_cHO1}
Q(z,t) = |\langle z |\hat{U}|\psi(0)\rangle|^2,    
\end{equation}
which may be re-written as 
\begin{equation}
\label{eqn:Qttranspz}
Q(z,t)= |\langle \epsilon(t)|\psi(0)\rangle|^2, 
\end{equation}
where we define
\begin{equation}
|\epsilon(t)\rangle = \hat{U}^{\dagger}|z\rangle.
\end{equation}
The time evolution operator $\hat{U}^{\dagger}$ can be interpreted as the time evolution generated by $-\hat K^\dagger$. In general, $|\epsilon(t)\rangle$ will not be a coherent state. In the classical limit, however, one can approximate $|\epsilon(t)\rangle$ with a coherent state and a norm factor as  \cite{graefe2010classical}
\begin{equation}
    |\epsilon(t)\rangle=w(z,t)^{\frac{1}{2}}|\tilde z(t)\rangle,
\end{equation}
where $\tilde z(t)$ is the solution of a complexified Hamiltonian flow with the Hamiltonian $-K^*$, and $w(z,t)^\frac{1}{2}$ the corresponding norm, yielding (\ref{eqn:Husimi_solution}). 

The quantum and classical Husimi flows are identical as long as the Hamiltonian is a bilinear function of the creation and annihilation operators. For more general systems the classical Husimi evolution (\ref{eqn:Husimi_solution}) is a short-time approximation of the full quantum dynamics.  We will illustrate the features of the classical Husimi flow and its relation to the quantum flow in two examples. 

\begin{figure}[t]
\hspace{-0.312cm}
\includegraphics[width=0.5\textwidth]{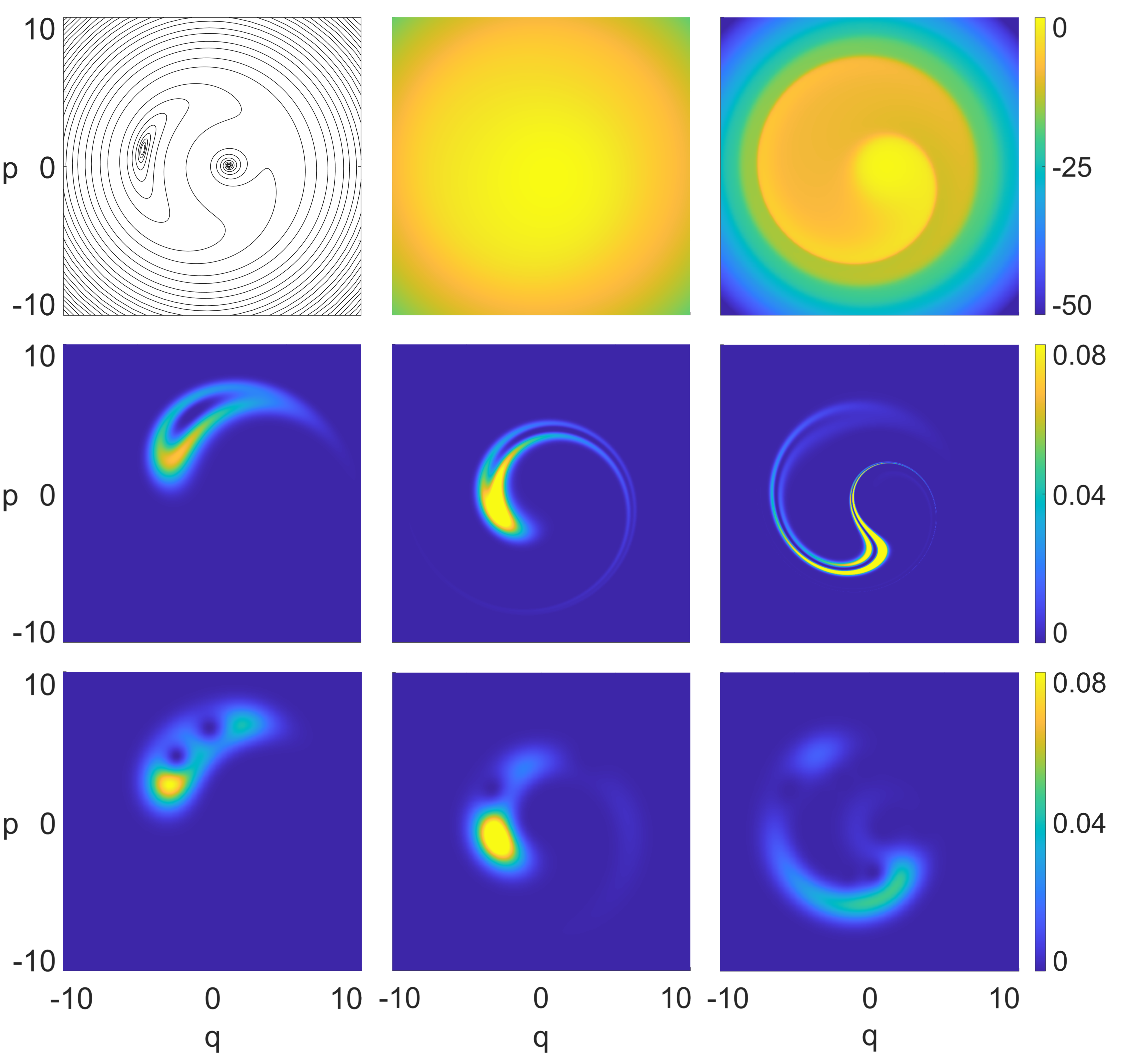}\hspace{-0.312cm}
\caption{\label{fig_mexican_hat} Non-Hermitian anharmonic oscillator (\ref{eqn:MH_Ham}) with $\gamma=0.05$, $\beta=0.05$, $\delta=1$. Top row: Phase-space trajectories (left) and logarithm of norm landscapes at $t=2$ (middle) and $t=8$ (right). Snapshots of the classical (middle row) and quantum (bottom row) Husimi distributions (renormalised) at times $t=0.5$ (left), $t=2$ (middle) and $t=8$ (right), for an initial harmonic oscillator state $|2\rangle$ displaced to $z_c=\frac{1}{\sqrt{2}}(-3+5i )$.}
\end{figure}

\textit{Example 1.}---As a first example we consider a tilted Mexican hat profile in phase space with additional damping, described by the   Hamiltonian
\begin{equation}
\label{eqn:MH_Ham}
  \hat K=-\hat a^\dagger \hat a-i\gamma\hat a^\dagger\hat a+\beta\hat a^\dagger\hat a^\dagger\hat a \hat a+\frac{\delta}{\sqrt{2}}(\hat a^\dagger+\hat a),
\end{equation}
with $\beta,\delta,\gamma\in\mathds{R}^+$. Similar models arise in various physical systems, such as optical systems with a Kerr nonlinearity \cite{drummond1980quantum,rigo1997quantum}. The anti-Hermitian part can be viewed as resulting from photon losses and a post selection of experimental outcomes \cite{klauck2019observation}. The classical Hamiltonian is given by
\begin{align}
      K=& -|z|^2-i\gamma|z|^2+\beta|z|^4+\frac{\delta}{\sqrt{2}}(z^*+z)\\
      \nonumber    =&-(1+i\gamma)\frac{p^2+q^2}{2}+\beta\frac{(p^2+q^2)^2}{4}+\delta q.
\end{align}
A phase-space portrait of the trajectories along which the Husimi distribution is transported is depicted in the top left panel in figure \ref{fig_mexican_hat}, for the parameter values $\beta=0.05$, $\delta=1$, and $\gamma=0.05$. The trajectories spiral outwards away from two fixed points. Two snapshots of the logarithm of the corresponding norm landscape are depicted in the middle and right plots of the top row of the same figure at times $t=2$ and $t=8$, respectively. We observe a dominant maximum located away from the origin. The overall norm decreases, and  the final value of the norm depends strongly on the initial positions. The remaining panels of figure \ref{fig_mexican_hat} shows snapshots of the Husimi distribution corresponding to an initially displaced second excited state. The middle row shows the classical flow, the bottom row shows the full quantum Husimi distribution at the same times.  We observe clear traces of the phase-space trajectories in the time evolved Husimi distribution for these intermediate times. The quantum flow is organised according to the underlying classical flow, with additional quantum interferences. The long term dynamics of both quantum and classical flow are dominated by the single maximum in the norm landscape.

The main contribution of the anti-Hermitian part in this example is a damping effect. This is a typical feature, but by no means the only possible effect of anti-Hermitian terms. In particular in $PT$-symmetric cases the interplay between Hermitian and anti-Hermitian components can lead to far less intuitive behaviours. We will close with a brief discussion of the Husimi flow in one such example. 

\textit{Example 2.}--- We consider the Hamiltonian
\begin{equation}\label{eqn:PT_Ham}
  \hat K=\hat a^\dagger \hat a+\beta\hat a^\dagger\hat a^\dagger\hat a \hat a-i\frac{\epsilon}{\sqrt{2}}(\hat a^\dagger+\hat a),
\end{equation}
with $\beta,\epsilon\in\mathds{R}^+$. An anharmonic oscillator with a linear gain-loss profile in $q$ direction. 
This system is $PT$-symmetric with respect to the $PT$ operator
\begin{equation}
    PT:\quad \hat q\to -\hat q,\quad \hat p\to\hat p,\quad i\to-i.
\end{equation}

The corresponding classical Hamiltonian is given by 
\begin{align}
 \nonumber K&=|z|^2+\beta|z|^4-i\frac{\epsilon}{\sqrt{2}}(z^*+z)\\
  &=\frac{p^2+q^2}{2}+\frac{\beta}{4}(p^2+q^2)^2-i\epsilon q.\label{eqn:classicalK_PTMil}
\end{align}
The resulting phase-space trajectories are equivalent to the trajectories of the real-valued Hamiltonian
\begin{equation}
\label{eqn:classicalH_PTMil}
  \tilde H=\frac{p^2+q^2}{2}+\frac{\beta}{4}(p^2+q^2)^2+\epsilon p.
\end{equation}
The imaginary part of the Hamiltonian (\ref{eqn:classicalK_PTMil}) leads to a pull towards decreasing position, which is reflected in the extra linear momentum term in the real Hamiltonian. There is a single elliptic fixed point located at $q=0$ and $p_0<0$, where $p_0$ is the single real root of the polynomial $p_{0}^3+\frac{1}{\beta}p_{0}+\frac{\epsilon}{\beta}=0$.

\begin{figure}[t]
\includegraphics[width=0.5\textwidth]{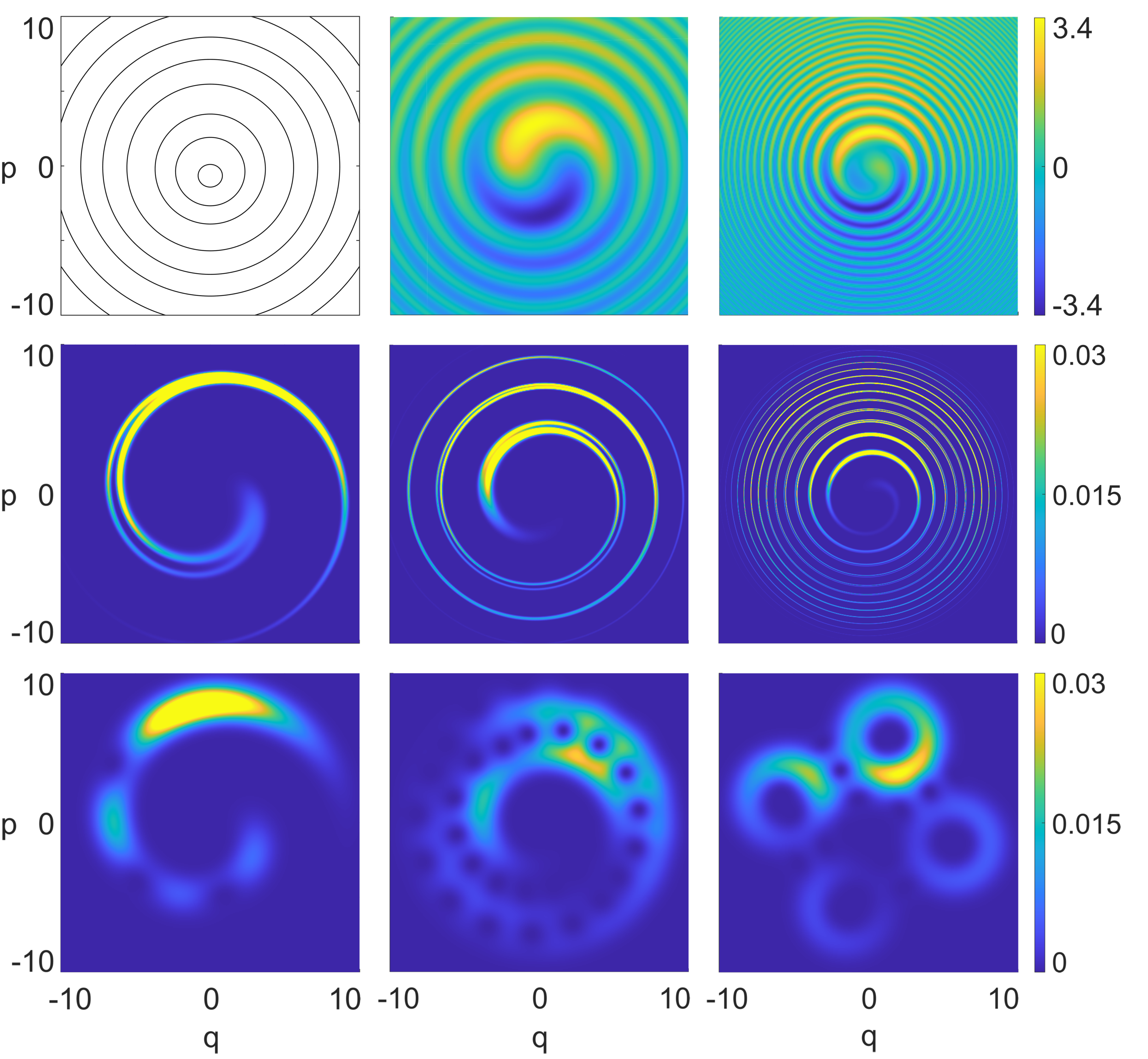}\hspace{-0.312cm}
\caption{\label{fig_PT_symm} As figure \ref{fig_mexican_hat}, for the non-Hermitian \textit{PT} symmetric system (\ref{eqn:PT_Ham}) with $\beta=0.25$ and $\epsilon=1$, and times $t=\frac{\pi}{10}$ (left), $t=\frac{\pi}{4}$(middle) and $t=\pi$ (right), for an initial harmonic oscillator state $|3\rangle$ displaced to $z_c=\frac{1}{\sqrt{2}}(5+3i )$.}

\end{figure}

A phase-space portrait for $\beta=0.25$ and $\epsilon=1$ is depicted in the top left panel in figure \ref{fig_PT_symm}. Due to the quartic term, the frequencies of the orbits increase with their distance from the fixed point. When transporting an initial density along these orbits the different frequencies lead to a spread of the initial distribution along whorl structures, similar to the Hermitian case \cite{milburn1986quantum}. The non-Hermiticity of the system manifests in the time-dependence of the norm landscape, two snapshots of which (at times $t=\frac{\pi}{4}$ and $t=\pi$) are depicted in the middle and right plots in the top row of the same figure. Since the trajectories are all symmetric around $q=0$ the value of the norm periodically returns to one for all initial states, however, at different times for the different initial conditions. In contrast to the previous example, the values of the classical norm are bounded, and the long time limit of the norm landscape is given by a distribution that is spread out over an ever finer spiralling structure. The middle and bottom rows of figure \ref{fig_PT_symm} show snapshots of the classical and quantum dynamics of the Husimi distribution for an initially displaced third excited state at three different times. The classical Husimi distribution evolves into a whorl structure, where the density is concentrated in regions corresponding to the higher areas of the norm landscape. The full quantum evolution shows characteristic interference effects on top of the classical flow, which are partially washed out by the non-Hermitian part. The non-uniform distribution of the quantum density is clearly in accordance with the predictions of the classical norm landscape. 

\textit{Conclusion}---In summary we have derived a classical counterpart of the Husimi flow generated by non-Hermitian Hamiltonians, which highlights the stark differences to Hermitian systems. The classical flow consists of two factors, the original Husimi distribution transported along phase-space trajectories, and a norm landscape. The interplay of both factors leads to intriguing dynamics, which often describe damping effects, such as in our first example, or more subtle effects of gain and loss, such as in our second ($PT$-symmetric) example. The investigation of the classical Husimi flow will be a powerful tool in the analysis of general non-Hermitian model systems. It is an interesting task to extend the approach presented here to master equations and stochastic Schr\"odinger equations appearing in other models of open quantum systems. 

\textit{Acknowledgements}---E.M.G. and S.M. acknowledge  support  from from the European Research Council (ERC) under the European Union's Horizon 2020 research and innovation program (grant agreement No 758453), E.M.G acknowledges support from the Royal Society (Grant. No. URF\textbackslash R\textbackslash 201034).


\begin{thebibliography}{33}
\bibitem [{com()}]{comment_Wigner}%
  \BibitemOpen
  \href@noop {} {}\bibinfo {note} {A similar approach using the Wigner function
  leads to the same dynamical equation in the semiclassical limit in the
  Hermitian case. In the presence of an anti-Hermitian term, the equation of
  motion for the Wigner function has higher order derivatives even for
  quadratic Hamiltonians that cannot be neglected if the Wigner function is
  sufficiently localised on phase-space. In contrast, in the Husimi function
  the influence of the anti-Hermitian term is captured by the first order
  derivatives.}\BibitemShut {Stop}%
\end{thebibliography}
\end{document}